\documentclass
[aps,prl,amsfonts,amssymb,twocolumn,amsmath,preprintnumbers,nofootinbib,floatfix,
showpacs,superscriptaddress]{revtex4-1}%
\usepackage[dvips]{graphics}
\usepackage{graphicx}
\usepackage{dcolumn}
\usepackage{bm}
\usepackage{amsmath}
\usepackage{amsfonts}
\usepackage{amssymb}
\usepackage{xcolor}
\usepackage{subfigure}
\usepackage{hyperref,hypcap}
\usepackage{braket}
\usepackage{commath}%
\providecommand{\U}[1]{\protect\rule{.1in}{.1in}}

\begin{document}

\title{Linear magnetoresistance induced by intra-scattering semiclassics of Bloch electrons}
\author{Cong Xiao}
\affiliation{Department of Physics, The University of Texas at Austin, Austin, Texas 78712, USA}
\author{Hua Chen}
\email{huachen@colostate.edu}
\affiliation{Department of Physics, Colorado State University, Fort Collins, CO 80523, USA}
\affiliation{School of Advanced Materials Discovery, Colorado State University, Fort Collins, CO 80523, USA}
\author{Yang Gao}
\author{Di Xiao}
\affiliation{Departiment of Physics, Carnegie Mellon University, Pittsburgh, PA 15213, USA}
\author{Allan H. MacDonald}
\author{Qian Niu}
\affiliation{Department of Physics, The University of Texas at Austin, Austin, Texas 78712, USA}

\begin{abstract}
The weak field magnetoresistance has seen a revived interest due to the
distinct role played by the momentum-space Berry curvature of Bloch electrons.
While most previous studies in this regard focus on the inter-scattering
motion of semiclassical Bloch electrons in electromagnetic fields, the
intra-scattering effects of the semiclassical dynamics augmented by the Berry
curvature, magnetic moment and shift vector on the magnetoresistance have been
largely overlooked. Here we uncover that these intra-scattering effects, which
are neglected in the field-independent relaxation time approximation to the
Boltzmann collision integral, can be as important as the inter-scattering
ones. Concrete calculations on the two dimensional gapped Dirac model show
that the sign of the negative linear magnetoresistance given by the Berry
curvature alone is reversed when one considers the magnetic moment and shift vector.
\end{abstract}
\maketitle


\emph{{\color{blue} Introduction}}--- The magnetoresistance (MR) of a
conductor is one of the central quantities in magnetotransport experiments
\cite{Ashcroft}. Distinct MR characteristics are associated with
superconductivity, weak localization, quantum Hall effect and other key
phenomena of condensed matter physics. The recent great interest in the MR in
the weak magnetic field regime stems from appreciating the importance of the
momentum space Berry curvature of Bloch electrons \cite{Xiao2010}. The
observed MR behaviors in exotic materials such as topological insulators and
topological semimetals are believed to be explained by the Berry curvature
modified semiclassical equations of motion \cite{Xiao2010} combined with the
Boltzmann transport equation in the relaxation time approximation
\cite{Armitage2018,Bernevig2018,Lu2017}. However, such a treatment
\cite{Son2013,Kim2014,Spivak2016,Du2017} merely includes the inter-scattering
motion of Bloch electrons driven by electromagnetic fields but dismisses the
effects of the semiclassical dynamics during scattering. The latter has proven
to be important in the linear \cite{Sinitsyn2008} and nonlinear
\cite{Konig2019,Du2019,Xiao2019,Fu2018,Sodemann2019} anomalous Hall effects,
even reversing the sign predicted by the Berry curvature contribution alone
\cite{Fu2015}. This naturally poses the question on the role of
intra-scattering semiclassics in MR.

\begin{figure}[h]
\begin{center}
\includegraphics[width=3 in]{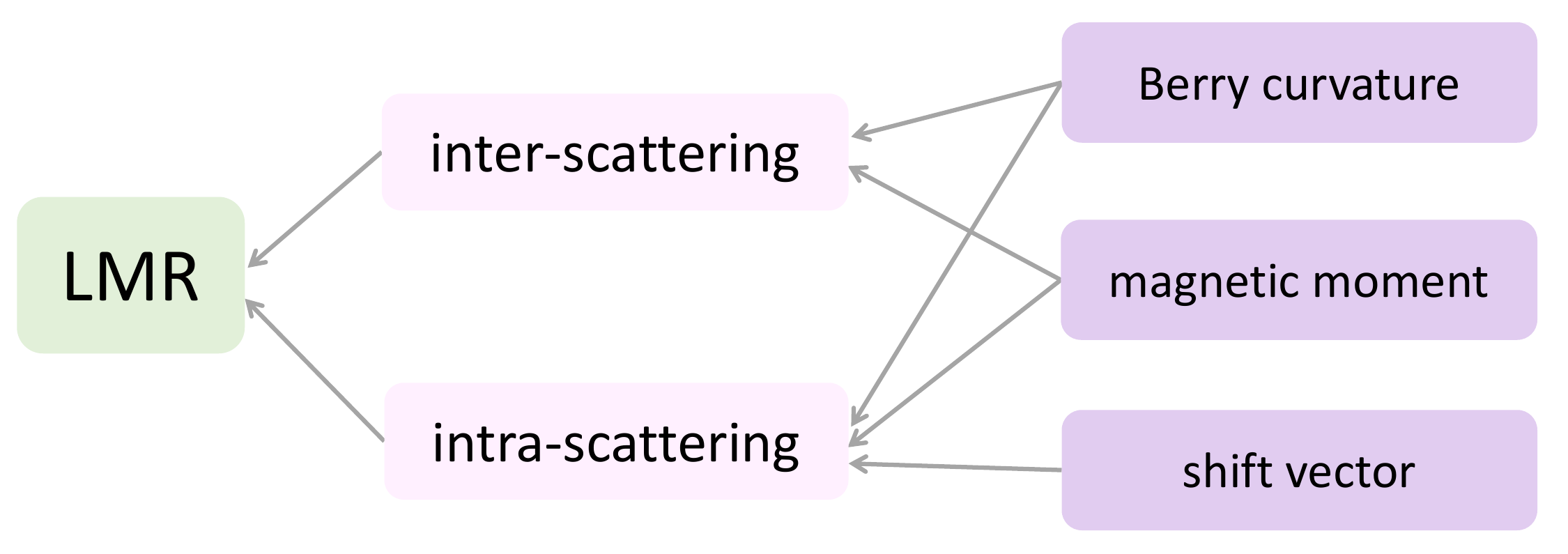}
\end{center}
\caption{(Color online) Mechanisms contributing to the LMR in the zeroth order
of scattering time in crystalline conductors with broken time-reversal
symmetry.}%
\label{fig:1}%
\end{figure}

In this Rapid Communication, we highlight the necessity of incorporating
intra-scattering effects of semiclassical dynamics of Bloch electrons,
embodied in geometrical quantities encoding quantum mechanical interband
coherence such as the Berry curvature, orbital magnetic moment \cite{Xiao2010}
and shift vector (or side jump) \cite{SJ1982,SJ1988,Sinitsyn2006}, into the
study of the MR (Fig.~\ref{fig:1}). In this article we focus on the linear
magnetoresistance (LMR) that is allowed in systems with broken time-reversal
symmetry \cite{Onsager,Gao}. We give explicit formulas for the various
contributions to the LMR in the first order of relaxation time $\tau$ (or
zeroth order for the ratio $\delta\rho/\rho$). Such LMR would break the
empirical Kohler's rule, which says that $\delta\rho/\rho$ depends on the
magnetic field only through $\omega_{c}\tau$, where $\omega_{c}$ is the
cyclotron frequency. As our theory is valid in the semiclassical regime
where $\omega_{c}\tau\ll1$, the $\tau^{0}$ contribution to
$\delta\rho/\rho$ identified in this work is expected to persist to
higher fields in more disordered systems and/or at elevated
temperatures for which $\tau$ is smaller. In this regime it dominates
over the $\tau^{1}$ contribution from the skew scattering, which is not
discussed here.

Previous studies on the LMR induced by Berry curvature related semiclassics
mainly concentrate on the 3D Weyl semimetals
\cite{Cortijo2016,Sharma2017,Zyuzin2017,Mertig2019}. In such systems the
anisotropic band structures and the presence of intervalley scattering have
hampered the analysis of intra-scattering processes. Only recently was the
Berry curvature induced momentum space measure \cite{Xiao2005} taken into
account in the Boltzmann collision integral \cite{Mertig2019}, but the roles
of the magnetic moment and shift vector still have not been addressed. To
bring out the essential physics we calculate the transverse MR in an isotropic
2D gapped Dirac model, which is a minimal model supporting such
intra-scattering contributions. Our calculations show that the contributions
from intra- and inter-scattering processes are of similar magnitude, and the
sign of the negative LMR contributed by the Berry curvature alone is reversed
when including the magnetic moment and shift vector. Because of the interplay
between inter- and intra-scattering contributions, the weak field
magnetotransport is much richer than usually assumed. In the end we will
discuss a few implications of our work on, e.g. the MR scaling law
\cite{Du2019} and MR in Weyl semimetals.

\emph{{\color{blue} Preliminaries}}--- In the presence of uniform, static
electric and magnetic fields the steady-state Boltzmann equation for the
semiclassical distribution function $f_{l}$ reads%
\begin{equation}
\frac{1}{D_{l}}\frac{e}{\hbar}\left(  \mathbf{E}+\mathbf{\tilde{v}}_{l}%
\times\mathbf{B}\right)  \cdot\frac{\partial f_{l}}{\partial\mathbf{k}}%
=-\sum_{l^{\prime}}D_{l^{\prime}}\omega_{ll^{\prime}}\left(  f_{l}%
-f_{l^{\prime}}\right)  , \label{SBE}%
\end{equation}
where $e=-|e|$, and $l=\left(  n,\mathbf{k}\right)  $ is the index for a Bloch
state in band $n$ with crystal momentum $\mathbf{k}$. The noncanonical dynamic
structure of semiclassical Bloch electrons shaped by the Berry curvature
$\mathbf{\Omega}_{l}$ implies a modified reciprocal space measure
\cite{Xiao2005} $D_{l}=1-\frac{e}{\hbar}\mathbf{B}\cdot\mathbf{\Omega}_{l}$.
Meanwhile, a semiclassical electron residing on a particular band is
physically identified as a finite-size wave-packet, and its self-rotation
gives rise to an intrinsic orbital magnetic moment $\mathbf{m}_{l}$. We also
absorb the spin magnetic moment of a Bloch wave-packet in $\mathbf{m}_{l}$
hereinbelow. The corresponding energy shift $-\mathbf{m}_{l}\cdot\mathbf{B}$
is independent of the cyclotron center-of-mass motion of the wave-packet
driven by Lorentz force, but contributes to the band velocity $\mathbf{\tilde
{v}}_{l}=\partial_{\hbar\mathbf{k}}\tilde{\epsilon}_{l}$, where $\tilde
{\epsilon}_{l}=\epsilon_{l}-\mathbf{B}\cdot\mathbf{m}_{l}$.

In the following we expand Eq.~\eqref{SBE} up to the first order in $E$ and
$B$, assuming weak fields. A quantity $X$ of the $i$-th order in $E$ and
$j$-th order in $B$ is denoted by a pair of superscripts: $X^{ij}$. Ignoring
the electric field for now and considering elastic scattering only, the Born
scattering rate reads $\omega_{ll^{\prime}}=\frac{2\pi}{\hbar}W_{ll^{\prime}%
}\delta\left(  \epsilon_{l}-\mathbf{m}_{l}\cdot\mathbf{B}-\epsilon_{l^{\prime
}}+\mathbf{m}_{l^{\prime}}\cdot\mathbf{B}\right)  $, where $W_{ll^{\prime}}$
is the scattering amplitude. Up to the linear order in $B$ we can obtain from
the above equation
\begin{align}
&  \omega_{ll^{\prime}}^{00}=\frac{2\pi}{\hbar}W_{ll^{\prime}}\delta\left(
\epsilon_{l}-\epsilon_{l^{\prime}}\right)  ,\\
&  \omega_{ll^{\prime}}^{01}=\frac{2\pi}{\hbar}W_{ll^{\prime}}\frac
{\partial\delta\left(  \epsilon_{l}-\epsilon_{l^{\prime}}\right)  }%
{\partial\epsilon_{l}}\mathbf{B}\cdot\left(  \mathbf{m}_{l^{\prime}%
}-\mathbf{m}_{l}\right)  .\nonumber
\end{align}

The contribution from $\mathbf{E}$ to $\omega_{ll^{\prime}}$ has to do with
the side jump \cite{Sinitsyn2006,SJ1982,SJ1988}
\begin{equation}
\omega_{ll^{\prime}}^{10}=\frac{2\pi}{\hbar}W_{ll^{\prime}}\frac
{\partial\delta\left(  \epsilon_{l}-\epsilon_{l^{\prime}}\right)  }%
{\partial\epsilon_{l}}e\mathbf{E}\cdot\delta\mathbf{r}_{l^{\prime},l}%
\end{equation}
where $\delta\mathbf{r}_{l^{\prime},l}=\mathbf{A}_{l^{\prime}}-\mathbf{A}%
_{l}-\left(  \partial_{\mathbf{k}^{\prime}}+\partial_{\mathbf{k}}\right)
\arg\langle u_{l^{\prime}}|u_{l}\rangle$ is the effective positional shift of
a semiclassical Bloch electron under elastic scattering. Here $\mathbf{A}_{l}$
is the Berry connection, $|u_{l}\rangle$\ is the periodic part of a Bloch
state, and $\arg$ means taking the argument of a complex number.

When both $\mathbf{E}$ and $\mathbf{B}$ are present we have
\begin{equation}
\omega_{ll^{\prime}}=\omega_{ll^{\prime}}^{00}+\omega_{ll^{\prime}}%
^{10}+\omega_{ll^{\prime}}^{01}. \label{scattering rate}%
\end{equation}
The $O(EB)$ term $\omega_{ll^{\prime}}^{11}$ is not needed since we
are interested in the electric current up to the order of $EB\tau$ \cite{note}.

Correspondingly, to solve the Boltzmann equation the distribution function can
be expanded in ascending powers of $\left(  E,B\right)  $ as
\begin{equation}
f_{l}=f_{l}^{00}+f_{l}^{10}+f_{l}^{01}+f_{l}^{11}+\dots
\end{equation}
where $f_{l}^{00}$ is the Fermi-Dirac distribution function without fields.
The details are relegated to the Supplemental Material \cite{Supp}.

\emph{{\color{blue} Absence of net equilibrium current}}--- The importance of
including $\mathbf{m}_{l}$ in the collision integral can already be
appreciated in the presence of $\mathbf{B}$ only. In $O\left(  E^{0}%
B^{1}\right)  $ the Boltzmann equation reads%
\begin{equation}
0=\mathcal{I}^{0}[f_{l}^{01}]+\mathcal{I}^{\text{B}}[f_{l}^{00}],
\end{equation}
where the functional
\begin{equation}
\mathcal{I}^{0}[g_{l}]\equiv\sum_{l^{\prime}}\omega_{ll^{\prime}}^{00}%
(g_{l}-g_{l^{\prime}})
\end{equation}
is the usual collision integral with $g_{l}$ being its argument, and%
\begin{equation}
\mathcal{I}^{\text{B}}[g_{l}]\equiv\sum_{l^{\prime}}\omega_{ll^{\prime}}%
^{01}(g_{l}-g_{l^{\prime}})
\end{equation}
is the correction to the collision functional due to the magnetic field.
Integration by parts yields
\begin{equation}
\mathcal{I}^{\text{B}}[f_{l}^{00}]=-\sum_{l^{\prime}}\omega_{ll^{\prime}}%
^{00}\mathbf{B}\cdot\left(  \mathbf{m}_{l^{\prime}}-\mathbf{m}_{l}\right)
\frac{\partial f_{l^{\prime}}^{00}}{\partial\epsilon_{l^{\prime}}},
\end{equation}
from which we get $f_{l}^{01}\sim\tau^{0}$:
\begin{align}
f_{l}^{01}=-\mathbf{B}\cdot\mathbf{m}_{l}\partial_{\epsilon_{l}}f_{l}^{00}.
\end{align}
This result is consistent with the usual employment of the equilibrium
distribution function $f^{\mathrm{eq}}\left(  \tilde{\epsilon}_{l}\right)
=f_{l}^{00}-\mathbf{B}\cdot\mathbf{m}_{l}\partial_{\epsilon_{l}}f_{l}^{00}$.

An important property of this $f^{\mathrm{eq}}$ is that the net current
vanishes in equilibrium, which can be verified by substituting the
semiclassical velocity containing the magnetic moment, $\mathbf{\dot{r}}%
=D_{l}^{-1}\mathbf{\tilde{v}}_{l}$, together with $f^{\mathrm{eq}}$ into the
equation for total current $\mathbf{j}=e\sum_{l}D_{l}\mathbf{\dot{r}%
}f^{\mathrm{eq}}_{l}=0$. Therefore, the collision integral $\mathcal{I}%
^{\text{B}}$ plays the key role in assuring that $f^{eq}\left(  \tilde
{\epsilon}_{l}\right)  $ is the solution of the Boltzmann equation in the
absence of electric field. In other words, the existence of the magnetic field
induced scattering rate $\omega_{ll^{\prime}}^{01}$ is necessary for the
validity of the Boltzmann equation in the presence of magnetic moment energy.

In previous studies on the MR, the effect of the magnetic moment on the
collision integral has been largely overlooked. This may be partially due to
the fact that in the absence of the electric field the magnetic field induced
scattering rate can be accounted for by simply promoting $f_{l}^{00}$ to
$f^{\mathrm{eq}}\left(  \tilde{\epsilon}_{l}\right)  $. However, in the
presence of an applied electric field, such a replacement alone without
considering the magnetic moment in the collision integral is inadequate. In
the following we illustrate that dismissing $\omega_{ll^{\prime}}^{01}$ would
result in the absence of a sizable contribution to the LMR in time-reversal
broken conductors.

\emph{{\color{blue} LMR}}--- To obtain the LMR of order $\tau^{0}$ that we are
interested in, the electric current in the order of $O\left(  EB\tau\right)  $
needs to be calculated. In $O\left(  EB\right)  $ the Boltzmann equation can
be grouped according to the scattering time dependence of the distribution
function, such as \cite{Supp}
\begin{equation}
f_{l}^{11}=f_{l}^{11,\left(  2\right)  }+f_{l}^{11,\left(  1\right)  }+....
\end{equation}
Here the superscript $\left(  i\right)  $ denotes the order of the scattering
time. Up to the linear order of $\tau$ we have the Boltzmann equation
$-\mathcal{I}_{c}^{\text{u}}[f_{l}^{11,\left(  2\right)  }]=e\left(
\mathbf{v}_{l}\times\mathbf{B}\right)  \cdot\partial_{\hbar\mathbf{k}}%
f_{l}^{10,\left(  1\right)  }$ for the usual Hall component $f_{l}^{11,\left(
2\right)  }$ and a Boltzmann equation responsible for $f_{l}^{11,\left(
1\right)  }$ \cite{Supp}:%
\begin{align}
-\mathcal{I}^{0}[f_{l}^{11,\left(  1\right)  }]  &  =e\mathbf{E}\cdot
\partial_{\hbar\mathbf{k}}f_{l}^{01}+2(D_{l}^{-1}-1)e\mathbf{E}\cdot
\partial_{\hbar\mathbf{k}}f_{l}^{00}\nonumber\\
&  +e\left(  \mathbf{v}_{l}\times\mathbf{B}\right)  \cdot\partial
_{\hbar\mathbf{k}}f_{l}^{10,\left(  0\right)  }+\mathcal{I}^{\mathrm{B}}%
[f_{l}^{10,\left(  1\right)  }]. \label{SBE-1}%
\end{align}
Here $f_{l}^{10}=f_{l}^{10,\left(  1\right)  }+f_{l}^{10,\left(  0\right)  }$
is the electric field induced out of equilibrium distribution function
determined by the $O(EB^{0})$ equation: $-\mathcal{I}^{0}[f_{l}^{10}%
]=e\mathbf{E}\cdot\partial_{\hbar\mathbf{k}}f_{l}^{00}+\mathcal{I}%
^{\mathrm{E}}[f_{l}^{00}]$, where the shift vector induced correction to the
collision integral reads
\begin{equation}
\mathcal{I}^{\mathrm{E}}[f_{l}^{00}]\equiv\sum_{l^{\prime}}\omega_{ll^{\prime
}}^{10}(f_{l}^{00}-f_{l^{\prime}}^{00})=-e\mathbf{E}\cdot\mathbf{v}%
_{l}^{\text{sj}}\partial_{\epsilon_{l}}f_{l}^{00},
\end{equation}
with $\mathbf{v}_{l}^{\text{sj}}=\sum_{l^{\prime}}\omega_{ll^{\prime}}%
^{00}\delta\mathbf{r}_{l^{\prime},l}$ being the side-jump velocity
\cite{Sinitsyn2006}.

The first term on the r.h.s. of Eq. (\ref{SBE-1}) arising from the magnetic
moment can be envisioned from $f^{\mathrm{eq}}\left(  \tilde{\epsilon}%
_{l}\right)  $ and has been considered before \cite{Du2017,Xiao2010}. When
writing down Eq. (\ref{SBE-1}) we have assumed that only one isotropic band on
the Fermi surface is involved in realistic scattering processes \cite{Supp}.
This is the case in some simplified model systems such as the 2D Dirac model,
or in the presence of long-range scattering potential for systems with
multiple locally isotropic valleys in the momentum space \cite{Xiao2010}. In
this case the factor 2 in the second term on the r.h.s. represents the effects
of Berry-curvature-corrected dynamics on both the driving and collision terms
of the Boltzmann equation (\ref{SBE}).

\begin{figure}[h]
\begin{center}
\includegraphics[width=2.5 in]{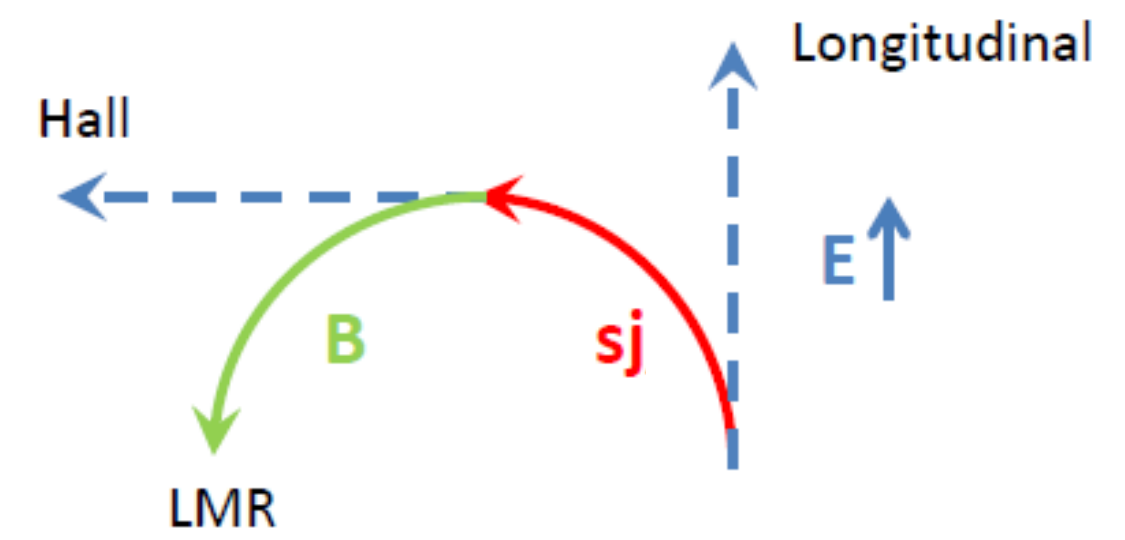}
\end{center}
\caption{(Color online) Schematics of the microscopic processes responsible
for the shift-vector (side-jump) contribution to the LMR.}%
\label{fig:2}%
\end{figure}

Now we turn to the second line of Eq. (\ref{SBE-1}) which is new. The first
term includes $f_{l}^{10,(0)}$ which is due to the side jump, and is
reminiscent of the side-jump contribution in the anomalous Hall effect. A
simple picture to understand the relation between the microscopic processes
leading to the LMR and to the anomalous Hall effect is illustrated in
Fig.~\ref{fig:2}. Consider two consecutive microscopic processes that can
individually give rise to a transverse component of the motion of a charge
carrier. One of them corresponds to the mechanisms that can lead to the
anomalous Hall effect, such as the side jump and skew scattering, and the
other is due to the Lorentz force. These two effects combined result in an
extra longitudinal component of the charge carrier's motion, and hence modify
the longitudinal conductance.

In contrast, the last term in Eq.~\eqref{SBE-1} has no counterpart in linear
responses. This term would have been omitted if the effect of magnetic moment
were accounted for only by promoting $f_{l}^{00}$ to $f^{\mathrm{eq}}\left(
\tilde{\epsilon}_{l}\right)  $. In the case of 2D isotropic bands with
long-range scattering, one has \cite{Supp}
\begin{equation}
\mathcal{I}^{\mathrm{B}}[f_{l}^{10,\left(  1\right)  }]=-B_{z}(\partial
_{\epsilon_{l}}m_{l})e\mathbf{E}\cdot\partial_{\hbar\mathbf{k}}f_{l}^{00},
\end{equation}
where the magnetic field is normal to the 2D plane and $\mathbf{m}_{l}%
=m_{l}\mathbf{\hat{z}}$. The corresponding off equilibrium distribution
function yields a current $e\sum_{l}\partial_{\hbar\mathbf{k}}\left(
-B_{z}m_{l}\right)  f_{l}^{10,\left(  1\right)  }$ along the electric field
direction and contributes to the linear MR. We note that in this special case
the contribution from $\mathcal{I}^{\mathrm{B}}[f_{l}^{10,\left(  1\right)
}]$ is the same as that from the magnetic moment induced velocity correction
of electrons.

Collecting above ingredients, we obtain
\begin{equation}
\mathbf{j}=e\sum_{l}[f_{l}^{11,\left(  2\right)  }\mathbf{v}_{l}^{\text{sj}%
}+f_{l}^{11,\left(  1\right)  }\partial_{\hbar\mathbf{k}}\epsilon_{l}%
+f_{l}^{10,\left(  1\right)  }\partial_{\hbar\mathbf{k}}\left(
-\mathbf{B\cdot m}_{l}\right)  ]
\end{equation}
in $O(EB\tau)$. The resulting LMR $\delta\rho/\rho\approx-\sigma_{xx}%
^{01}/\sigma_{xx}^{00}$ ($\sigma_{xx}$ is the longitudinal conductivity) is
independent of $\tau$ and breaks the Kohler's rule.

\begin{figure}[h]
\begin{center}
\includegraphics[width=2.9 in]{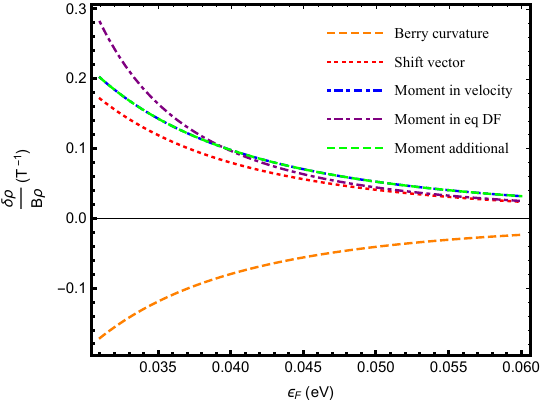}
\end{center}
\caption{(Color online) LMR in the 2D gapped Dirac model (\ref{model}). $\hbar
v_{F}=4.1$ eV$\cdot\mathrm{\mathring{A}}$ for Bi$_{2}$Se$_{3}$ according to
\cite{Zhang2009}. $h_{0}$ is set to be 20 meV. The $g$ factor for the spin
part of the magnetic moment is taken to be 50 \cite{Fisher2010}. The purple
line results from promoting the equilibrium distribution function (eq DF)
$f_{l}^{00}$ into $f^{eq}\left(  \tilde{\epsilon}_{l}\right)  $, whereas the
green line follows from $\mathcal{I}_{c}^{\text{m}}[f_{l}^{10,\left(
1\right)  }]$.}%
\label{fig:3}%
\end{figure}

\begin{figure}[h]
\begin{center}
\includegraphics[width=2.5 in]{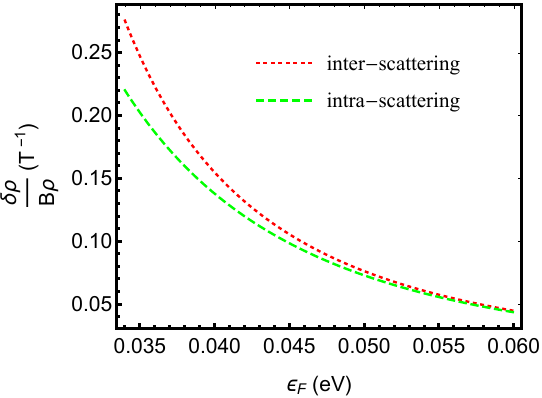}
\end{center}
\caption{(Color online) LMR in model (\ref{model}) arising from the
inter-scattering and intra-scattering processes. The contribution represented
by the purple line in Fig.~\ref{fig:3} is assigned to the inter-scattering
part, as it can be obtained by a simple promotion of the equilibrium
distribution function. }%
\label{fig:4}%
\end{figure}

\emph{{\color{blue} Model illustration}}--- Below we calculate the various
contributions to the LMR in $O(\tau^{0})$ in a minimal model -- the 2D gapped
Dirac model%
\begin{equation}
H=\hbar v_{F}\left(  \sigma_{x}k_{x}+\sigma_{y}k_{y}\right)  +h_{0}\sigma
_{z},\label{model}%
\end{equation}
where $\sigma_{i}$ are the Pauli matrices on the spin space. A spin magnetic
moment $g\mu_{B}\sigma_{z}/2$\ with the $g$-factor $g$ and Bohr magneton
$\mu_{B}$ can also be included in addition to the orbital moment. The Fermi
level $\epsilon_{F}$ is set to intersect the upper band. We assume the
longitudinal conductivity is much larger than the Hall one, hence
$\epsilon_{F}$ is not close to the conduction band bottom.
The results are presented in Fig.~\ref{fig:3}. Contributions from different
mechanisms are all of similar magnitude. Notably, taking into account the
magnetic moment and shift vector reverses the sign of the MR contributed by
the Berry curvature alone. The importance of the intra-scattering contribution
to LMR, which has the similar size to the inter-scattering one, is
demonstrated in Fig. \ref{fig:4}. This result indicates the inadequacy of the
naive relaxation time approximation in magnetotransport even with the
inter-scattering contribution properly accounted for.

\emph{{\color{blue} Discussion}}---In this section we discuss a number of
implications of our work on the theory of magnetotransport, which may be worth
exploring in the future.

The LMR studied in this work is independent of the transport relaxation time
only if there is a single type of disorder. Similar to the case of the linear
and nonlinear anomalous Hall effects \cite{Du2019,Hou2015,Mak2019}, in the
presence of multiple scattering sources such as phonons and interface
roughness, the LMR should change as the relative ratio of these competing
scatterings varies. A scaling relation, which links the LMR to the zero-field
longitudinal conductivity and takes a similar form as those proposed in
anomalous Hall effects \cite{Du2019,Hou2015,Mak2019}, can thus be anticipated
in varying-temperature experiments \cite{Mak2019} on disordered samples
\cite{Xiao2019scaling}, which can help to elucidate the involved mechanisms.

The valley-contrasting nature of the orbital magnetic moment in inversion
asymmetric 2D gapped graphene and 3D Weyl systems \cite{Xiao2007} is expected
to amplify the influence of the orbital moments in the collision integral on
magnetotransport, provided that the intervalley scattering is strong. This is
based on the fact that the orbital moment acquires the maximum change upon
intervalley scattering.

Despite that our model consideration focuses on the transverse MR in 2D
systems for simplicity, the revealed importance of the magnetic moment should
also be anticipated in the longitudinal MR in 3D systems. In particular, the
chiral anomaly related longitudinal negative MR in 3D Weyl semimetals arises
from the Berry curvature alone \cite{Son2013}, hence it is interesting to
inspect the interplay of the chiral anomaly, orbital moment and shift vector.
A few recent studies have included partly the inter-scattering effects of
orbital moments and found the reduction of the chiral anomaly negative MR
\cite{Cortijo2016,Zyuzin2017}. More systematic studies on the effect of
intra-scattering semiclassics are needed in the future to see if a sign change
similar to that in Fig. \ref{fig:3} is possible in the longitudinal MR of Weyl semimetals.

In this study we put emphasis on the effect of the magnetic field through its
coupling to the magnetic moment and Berry curvature in the collision integral.
However, the magnetic field can also enter the collision integral through the
scattering amplitude $W_{ll^{\prime}}$, namely, through modifying the Bloch
wave functions. While the spin part is easy to obtain by a Zeeman coupling
term in the original Hamiltonian \cite{Du2017,Brinkman2019}, the orbital part
from the minimal coupling is still waiting to be addressed in the weak
magnetic field case. In a 2D electron gas confined in a quantum well, the
orbital effect of an in-plane magnetic field on the scattering amplitude can
be addressed by the usual quantum mechanical perturbation theory, implying the
mixing of states from different quantum-confined subbands \cite{Tarasenko2008}%
. While such a perturbation treatment does not apply directly to perpendicular
magnetic fields in the 2D case, incorporating the field-corrected wave-packet
states of electrons \cite{Gao2015} into the analysis of Boltzmann collision
integral seems to be promising. Alternatively, a quantum kinetic theory
accommodating not only the Berry curvature \cite{Akihiko2017,Akihiko2018} but
also the magnetic moment and shift vector may also be developed to verify the
predictions made by the semiclassical Boltzmann approach.

Finally we note that, although some of the microscopic processes leading to
the LMR are closely related to those responsible for the anomalous Hall
effect, it is not necessarily true that these two phenomena appear together.
As the anomalous Hall conductivity is a time-reversal-odd pseudovector defined
as $\sigma_{\alpha}^{\mathrm{AH}}=\epsilon_{\alpha\beta\gamma}\sigma
_{\beta\gamma}/2$, where $\epsilon_{\alpha\beta\gamma}$ is the Levi-Civita
symbol and $\sigma_{\beta\gamma}$ is the conductivity tensor, its existence
requires such a pseudovector to be invariant under the operations in the
symmetry group of the ground state of the system. The low-field linear
magnetoconductivity, on the other hand, is part of a rank-3 pseudo tensor
$\sigma_{\alpha\alpha}^{01}=\chi_{\alpha\alpha\gamma}B_{\gamma}$, where the
repeated index $\alpha$ is not summed over. Assuming that the magnetic
structure is not distorted by the magnetic field in the weak field limit,
standard symmetry analysis of tensor properties \cite{Book} can be employed to
see if there should be linear MR in a given system. We find that the symmetry
requirements for a nonzero linear MR in magnetic systems are less stringent
than for the anomalous Hall effect. For example, in magnetic structures
possessing $D_{3}$, $C_{3v}$, or $D_{3d}$ symmetries, the anomalous Hall
effect is forbidden, but the linear MR can be nonzero when both the current
and the magnetic field are perpendicular to the 3-fold axis.

\emph{{\color{blue} Note added.}}--- During the completion of the present work
we were aware of a relevant paper \cite{Meng2019}, which re-examines the
longitudinal MR in inversion asymmetric Weyl semimetals by taking into account
the orbital magnetic moment in both intra- and inter-scattering processes. As
speculated in our discussion above, in Ref. \cite{Meng2019} the sign of the
longitudinal MR given by the Berry curvature mechanism alone is reversed if
the intervalley scattering is strong.

\begin{acknowledgments}
The authors are grateful to Cui-Zu Chang, Dan Ralph, Yong Chen, Fan Zhang, Yejun Feng, Yishu Wang and Tianlei Chai for helpful discussions. H.C. and A.H.M. were supported by ONR-N00014-14-1-0330 and Welch Foundation grant TBF1473.
Q.N. was supported by DOE (DE-FG03-02ER45958, Division of Materials
Science and Engineering) on the geometric formulation in this work.
C.X. and Y.G. was supported by NSF (EFMA-1641101) and Welch Foundation
(F-1255). D.X. was supported by AFOSR No. FA9550-14-1-0277.
\end{acknowledgments}


\begin{thebibliography}{99}                                                                                               %


\bibitem {Ashcroft}N. W. Ashcroft and N. D. Mermin, {\itshape}Solid State
Physics (Saunders, Philadelphia, 1976).

\bibitem {Xiao2010}D. Xiao, M.-C. Chang, and Q. Niu, Rev. Mod. Phys.
\textbf{82}, 1959 (2010).

\bibitem {Armitage2018}N. P. Armitage, E. J. Mele, and A. Vishwanath, Rev.
Mod. Phys. \textbf{90}, 015001 (2018).

\bibitem {Bernevig2018}A. Bernevig, H. Weng, Z. Fang, and X. Dai, J. Phys.
Soc. Jpn. \textbf{87}, 041001 (2018).

\bibitem {Lu2017}H.-Z. Lu and S.-Q. Shen, Front. Phys. \textbf{12}, 127201
(2017); H.-P. Sun and H.-Z. Lu, Front. Phys. \textbf{14}, 33405 (2019).

\bibitem {Son2013}D. T. Son and B. Z. Spivak, Phys. Rev. B \textbf{88}, 104412 (2013).

\bibitem {Kim2014}K.-S. Kim, H.-J. Kim, and M. Sasaki, Phys. Rev. B
\textbf{89}, 195137 (2014).

\bibitem {Spivak2016}B. Z. Spivak and A. V. Andreev, Phys. Rev. B \textbf{93},
085107 (2016).

\bibitem {Du2017}X. Dai, Z. Z. Du, and H.-Z. Lu, Phys. Rev. Lett.
\textbf{119}, 166601 (2017).

\bibitem {Sinitsyn2008}N. A. Sinitsyn, J. Phys.: Condens. Matter \textbf{20},
023201 (2008).

\bibitem {Konig2019}E. J. Konig, M. Dzero, A. Levchenko, and D. A. Pesin,
Phys. Rev. B \textbf{99}, 155404 (2019).

\bibitem {Fu2018}H. Isobe, S. Xu, and L. Fu, arXiv:1812.08162.

\bibitem {Du2019}Z. Z. Du, C. M. Wang, H.-Z. Lu, and X. C. Xie, Nat. Commun.
\textbf{10}, 3047 (2019).

\bibitem {Sodemann2019}S. Nandy and I. Sodemann, Phys. Rev. B \textbf{100},
195117 (2019).

\bibitem {Xiao2019}C. Xiao, Z. Z. Du, and Q. Niu, Phys. Rev. B \textbf{100},
165422 (2019).

\bibitem {Fu2015}I. Sodemann and L. Fu, Phys. Rev. Lett. \textbf{115}, 216806 (2015).

\bibitem {SJ1982}V. I. Belinicher, E. L. Ivchenko, and B. I. Sturman, Sov.
Phys. JETP \textbf{56}, 359 (1982).

\bibitem {SJ1988}Yu. B. Lyanda-Geller, Ferroelectrics \textbf{83}, 35 (1988).

\bibitem {Sinitsyn2006}N. A. Sinitsyn, Q. Niu, and A. H. MacDonald, Phys. Rev.
B \textbf{73}, 075318 (2006).

\bibitem {Onsager}In the weak field regime Onsager relations forbid any odd
powers of $B$ in the series expansion of the longitudinal conductivity
$\sigma_{xx}$ when only the magnetic field breaks time reversal symmetry (L.
Onsager, Phys. Rev. \textbf{37}, 405 (1931)). If time-reversal symmetry is
broken spontaneously, however, as in the case of a ferromagnet with
magnetization $\bm M$, Onsager relations require only that $\sigma_{xx}(-\bm
B,-\bm M)=\sigma_{xx}(\bm B,\bm M)$ and the linear MR is allowed.

\bibitem {Gao}The quadratic MR in time-reversal invariant systems is a
higher-order nonlinear response that requires field-corrected geometrical
quantities in both inter-scattering (Y. Gao, S. A. Yang, and Q. Niu, Phys.
Rev. B \textbf{95}, 165135 (2017).) and intra-scattering processes.

\bibitem {Cortijo2016}A. Cortijo, Phys. Rev. B \textbf{94}, 241105(R) (2016).

\bibitem {Sharma2017}G. Sharma, P. Goswami, and S. Tewari, Phys. Rev. B
\textbf{96}, 045112 (2017).

\bibitem {Zyuzin2017}V. A. Zyuzin, Phys. Rev. B \textbf{95}, 245128 (2017).

\bibitem {Mertig2019}A. Johansson, J. Henk, and I. Mertig, Phys. Rev. B
\textbf{99}, 075114 (2019).

\bibitem {Xiao2005}D. Xiao, J. Shi, and Q. Niu, Phys. Rev. Lett. \textbf{95},
137204 (2005).

\bibitem {note}$\omega_{ll^{\prime}}^{11}$ only contributes to conductivities
in the zeroth order of $\tau$. In time-reversal broken metals this
contribution is expected to be overshadowed by the O($\tau^{1}$) terms.

\bibitem {Supp}See Supplemental Material for the details of the order by order
analysis of the Boltzmann equation.

\bibitem {Zhang2009}H. Zhang, C.-X. Liu, X.-L. Qi, X. Dai, Z. Fang, S.-C.
Zhang, Nature Phys. \textbf{5}, 438 (2009).

\bibitem {Fisher2010}J. G. Analytis, R. D. McDonald, S. C. Riggs, J.-H Chu, G.
S. Boebinger, and I. R. Fisher, Nature Phys. \textbf{6}, 960 (2010).

\bibitem {Hou2015}D. Hou, G. Su, Y. Tian, X. Jin, S. A. Yang, and Q. Niu,
Phys. Rev. Lett. \textbf{114}, 217203 (2015).

\bibitem {Mak2019}K. Kang, T. Li, E. Sohn, J. Shan, and K. F. Mak, Nat. Mater.
\textbf{18}, 324-328 (2019).

\bibitem {Xiao2019scaling}C. Xiao, H. Zhou, and Q. Niu, Phys. Rev. B
\textbf{100}, 161403(R) (2019).

\bibitem {Xiao2007}D. Xiao, W. Yao, and Q. Niu, Phys. Rev. Lett. \textbf{99},
236809 (2007).

\bibitem {Brinkman2019}J. C. de Boer, D. P Leusink, and A. Brinkman, J. Phys.
Commun. \textbf{3,} 115021 (2019).

\bibitem {Tarasenko2008}S. A. Tarasenko, Phys. Rev. B \textbf{77}, 085328 (2008).

\bibitem {Gao2015}Y. Gao, S. A. Yang, and Q. Niu, Phys. Rev. B \textbf{91},
214405 (2015).

\bibitem {Akihiko2017}A. Sekine, D. Culcer, and A. H. MacDonald, Phys. Rev. B
\textbf{96}, 235134 (2017).

\bibitem {Akihiko2018}A. Sekine and A. H. MacDonald, Phys. Rev. B \textbf{97},
201301(R) (2018).

\bibitem {Book}R. R. Birss, {\itshape}Symmetry and Magnetism (North-Holland,
Amsterdam, 1966).

\bibitem {Meng2019}A. Knoll, C. Timm, and T. Meng, arXiv: 1912.07852
\end{thebibliography}
\end{document}